\documentstyle[12pt]{article}
\begin{document}

\title {Combined constraints on the SUSY parameter space from
        ~$\Delta r$ ~and Higgs boson search
\thanks{Supported in part by the Polish Committee for Scientific
        Research under the grant 2 0165 91 01
        and European Union under contract CHRX-CT92-0004}}
\author{Piotr H. Chankowski
\thanks{On leave of absence from
Institute of Theoretical Physics, Warsaw University.}\\
Institut f\"ur Theoretische Physik, Universit\"at Karlsruhe\\
Kaiserstrasse 12, 76128 Karlsruhe, Germany\\
and\\
Istituto Nazionale di Fisica Nucleare, Sezione di Padova\\
via F. Marzolo 8, 35131 Padova, Italy.
}

\maketitle

\vspace{-12cm}
\begin{flushright}
KA-TP 1994-4\\
\end{flushright}
\vspace{12cm}

\begin{abstract}
Combining the constraints coming from the ~$M_W$ ~measurements and the
unsuccesful search for the Higgs boson at LEP we determine in the framework
of MSSM the allowed mass regions for the lighter scalar partner of the top
quark. For a heavy top quark
particularily strong bounds are obtained for low values of ~$\tan\beta\equiv
v_2/v_1$ ~and light bottom squark.
\end{abstract}

\newpage

Recent calculation of the quantity ~$\Delta r$, ~i.e. of the ~$W^{\pm}$ ~boson
mass in terms of ~$G_F$, ~$\alpha_{EM}$ ~and ~$M_Z$ ~$-$ the three at present
best measured electroweak observables, within the Minimal Supersymmetric
Standard Model (MSSM) revealed \cite{MY_DR,DS} that, for fixed ~
$m_t$ $-$ the top quark mass, it is equal to or lower than ~$\Delta r$ ~in
the Standard Model (SM) with light, ~${\cal O}(M_Z)$, ~Higgs boson~$\phi^0$.
There are two reasons for such behaviur of ~$\Delta r$ ~in MSSM.
Firstly, all genuine SUSY particles (like squarks, sleptons, charginos) give
negative contributions to ~$\Delta r$. ~Therefore in order to maximize it
one has to keep those particles very heavy, say ~${\cal O}$(1-2 TeV).
Their contribution to ~$\Delta r$ ~is then negligibly small as
follows from the Appelquist -- Carazzone decoupling theorem. Secondly,
as is well known \cite{HHG}, the Higgs sectors of MSSM and SM are diferent.
All the masses and couplings of five MSSM Higgs bosons, ~$h^0$, ~$H^0$, ~
$A^0$ ~and ~$H^{\pm}$, ~can be (at the tree level) parametrized by two
variables only: ~$M_{A^0}$ - the ~$CP-$odd Higgs boson mass and ~$\tan\beta$
$-$the ratio of the two vacuum expectation values of the two Higgs
doublets. For given ~$\tan\beta$, ~maximum of ~$\Delta r$ ~is attained for ~
$M_{A^0}\gg M_Z$. ~In this limit, however, ~$A^0$, ~$H^0$ ~and ~$H^{\pm}$ ~
effectively decouple and the Higgs sector of MSSM is SM --like with one
light Higgs boson ~$h^0$. For large ~$\tan\beta$, which maximize ~$\Delta r$, ~
$M_{h^0}\sim {\cal O}(M_Z)$ ~thus
explaining why the upper limit for ~$\Delta r$ ~in MSSM does not exceed the
SM value of ~$\Delta r$ ~for ~$M_{\phi^0}\sim M_Z$.

When the SUSY particle masses are in the 100 GeV range their negative
contribution to ~$\Delta r$ ~may be substantial.
Two major sources of such negative contributions have been identified. Both
are related to rather large violation of the so called custodial ~$SU_V(2)$ ~
symmetry by sparticle masses (much in the same way as it is violated by the
top $-$ bottom mass splitting leading to decrease of ~$\Delta r$ ~as ~$m_t$ ~
grows). One is the slepton sector with sneutrinos lighter than ~$M_Z$.
In such a case ~$\Delta r$ ~receives large (larger for larger ~$\tan\beta$)
negative contribution due to ~$SU_V(2)$ ~violating sneutrino $-$left handed
charged slepton mass spliting. The second source is the similar mass spliting
between stop and sbottom masses.
The second effect, which is bigger for smaller values of ~$\tan\beta$, ~is
more interesting because it grows with ~$m_t$ which enters the formulae for
the top squark masses.

With the new, improved, data: ~$M_Z = 91.1895\pm0.0044$ GeV ~\cite{BP}
and ~$M_W = 80.23\pm0.18$ GeV ~\cite{MW}, the quantity ~$\Delta r$
\begin{eqnarray}
\Delta r = 1 - {\pi\alpha_{EM}\over\sqrt{2} G_F}
           {1\over M^2_W\left(1 - {M^2_W\over M^2_Z}\right)}
\end{eqnarray}
is constrained to ~$\Delta r = 0.044\pm0.010$ (notice the change as
compared with \cite{MY_DR} where ~$\Delta r = 0.044\pm0.015$ which mainly
reflects the improvement in ~$M_W$ ~measurement).
This means that at ~1$\sigma$ ~level
the top quark mass in MSSM is constrained to ~$m_t < 177$ GeV. Though
the 1$\sigma$ bound on ~$\Delta r$ ~may prove too restrictive, it is clear
that for a heavy top quark very little room
is left for acceptable negative contributions from squarks and/or sleptons.

This becomes particularily intriguing in view of the apparent discrepancy
between the value of the top mass ~$174\pm17$ GeV ~reported recently by the
CDF collaboration \cite{CDF_T} and the measurement of ~
$R_b\equiv\Gamma_{Z\rightarrow\bar bb}/\Gamma_{Z\rightarrow hadrons}$ ~at
LEP \cite{BP} which deviates from its Standard Model prediction
\cite{RB_SM} with ~$m_t=174$ GeV ~by  ~2$\sigma$. ~Such
discrepancy may well be due to the existence of a light SUSY partner of
the top quark \cite{BF}. In this context any constraint on the top squark
mass coming from quantities other than ~$R_b$ ~are very important.

In this note we want to explore more carefully such
constraints coming from  ~$\Delta r$ ~and the fact that no
SUSY scalar Higgs boson with mass less than roughly
50 GeV has been found so far \cite{LEP_H}. From the analysis of ~$\Delta r$ ~in
MSSM \cite{MY_DR} it follows that restricting ~$\Delta r$ ~to its
experimental ~$1\sigma$ ~bounds would implay that either SUSY
particles, in particular, squarks, are very heavy so that they are not
relevant for the
LEP1 physics or that the top quark mass is smaller than the central
value qiven by CDF measurement. Therefore, in what follows, as a useful
guideline we will consider two values of ~$m_t$, ~174 ~and ~160 GeV ~
together with the more conservative, ~2$\sigma$ ~bound on ~$\Delta r$ ~for
the former and ~1$\sigma$ ~bound for the latter.
Of course, the same analysis can be applied to any other
values of ~$m_t$ ~and/or more/less restrictive bounds on ~$\Delta r$.

The second constraint, ~$M_{h^0}>50$ GeV, becomes relevant because in
MSSM  quarks and squarks  have important
impact on the masses of the scalar Higgs bosons \cite{HJEBM,EZ}.
Dominant corrections are induced by the top and stop loops and are large
in the case of ~$m_t\gg M_W$. ~The general top
squarks mass squared matrix has the form:
\begin{eqnarray}
{\cal M}^2_{\tilde t_1,\tilde t_2} =
\left(\matrix{ M^2_{\tilde t_L} & -m_t A_t \cr
                    -m_t A_t & M^2_{\tilde t_R}}\right)
\end{eqnarray}
where
\begin{eqnarray}
M^2_{\tilde t_L} = m^2_Q + m_t^2 - {1\over6}\cos2\beta~(M_Z^2 - 4M^2_W)
v\nonumber\\
M^2_{\tilde t_R} = m^2_T + m_t^2 + {2\over3}\cos2\beta~(M_Z^2 - M_W^2)
\end{eqnarray}
Its two eigenvalues ~$M^2_{\tilde t_1}$ ~and ~$M^2_{\tilde t_2}$ ~are the
physical ~$\tilde t_1$ ~(lighter) and ~$\tilde t_2$ ~(heavier) top squark
masses. As follows from the simple estimate
\begin{eqnarray}
\Delta M^2_h\sim
{m_t^4\over M_W^2}
\log\left({M_{\tilde t_1}M_{\tilde t_2}\over m_t^2}\right) + ...
\end{eqnarray}
in contrast with the case of top squarks
heavier than top quark considered previously \cite{HJEBM,EZ,MY_H}, for top
squarks lighter than the top quark the corrections to the Higgs
boson masses can be negative.
The complete formulae for the corrections to
the Higgs boson masses in the, so called, effective potential approximation
can be found in e.g. \cite{EZ}. For our purpose this method is accurate enough
as follows from the detailed comparison of its results with the full
diagramatic 1--loop calculations \cite{MY_MH}  the error of the approximation
being smaller than 3$-$5 GeV.

{}For low values of ~$\tan\beta$, ~for which
the tree level lighter Higgs boson mass ~$M^{tree}_{h^0}$ ~is small
(it is bounded from above by ~$M^{tree}_{h^0} < |\cos2\beta |M_Z$ ~and
therefore is smaller than ~50 GeV ~for ~$\tan\beta < 1.85$) ~the
positive radiative corrections are crucial for making such small
values of ~$\tan\beta$ ~still phenomenologically acceptable.
Therefore, for given ~$m_t$ ~the requirement ~$M_{h^0}>50$ GeV ~puts some
restrictions on the top squark masses.

The elipses in eq.(4) stand for other terms which, in particular, tend
to decrease ~$\Delta M^2_{h^0}$ ~in presence of large
{\sl left--right} mixing in the top squarks sector i.e. when ~
$m_t A_t\sim M^2_{\tilde t_{L,R}}$. ~This is very important in view
of the fact that large {\sl left--right} mixing decreases the absolute
value of the negative
contribution of light squarks to ~$\Delta r$ ~and therefore seems to be
necessary for making ~$\Delta r$ ~acceptable.

If ~$A_t=0$, ~i.e. in the case of no {\sl left-right} mixing in the stop
sector (mixing in the sbottom and other sectors can be neglected if naturality
is invoked) mass eigenstates are simply ~$\tilde t_L$ ~and ~$\tilde t_R$ ~
with ~$\tilde t_1 \equiv \tilde t_L$ ~or ~$\tilde t_R$ ~depending on which
one is lighter. ~$M_{\tilde t_L}$ ~is related to ~$M_{\tilde b_L}$ ~given by
\begin{eqnarray}
M_{\tilde b_L} = m^2_Q + m^2_b - {1\over6}\cos2\beta~(M^2_Z + 2M^2_W)
\end{eqnarray}
since they contain one common free soft SUSY breaking parameter ~$m^2_Q$.
Therefore the lower bound on the bottom squark mass of 120 GeV
puts also lower bound on the mass of purely left handed
stop which depends on the top mass and ~$\tan\beta$. In general, because ~
$m_t > M_W$ ~we always have ~$M_{\tilde t_L} > M_{\tilde b_L}$ ~and this
mass splitting is bigger for smaller values of ~$\tan\beta$.

To demonstrate the importance of the Higgs boson mass bound we plot in Fig.1
the lower limit for the mass of the lighter top squark ~$M_{\tilde t_1}$ ~as
a function of the mass of the left--handed sbottom (we
neglect the small {\sl left-right} mixing in the bottom squark sector) for
few values of ~$\tan\beta$ ~and two values of the top quark mass. For
fixed ~$M_{\tilde b_L}$ ~a scan is performed over ~
$M_{\tilde t_R}$ ~and ~$A_t$ ~and the lowest value of ~$M_{\tilde t_1}$ ~
compatible with the requirement ~$M_{h^0} > 50$ GeV is plotted. We see that
particularily strong bounds are obtained for ~$1 < \tan\beta <2$. ~For ~
$\tan\beta > 2.5$ ~the constraint quickly dissapears. This is because
the tree level part of ~$M_{h^0}$ ~grows with ~$\tan\beta$ (for ~
$M_{A^0} \gg M_Z$ ~we have ~$M_{h^0} = |\cos2\beta| M_Z$) and therefore
there is more room for negative contribution to ~$M_{h^0}$ ~from radiative
corrections. The maxima seen in Fig.1 for ~$\tan\beta=$1.2 and 1.5 ~arise
roughly at the points
where ~$M_{\tilde t_R}^{min}$, ~for which the minimum of ~$M_{\tilde t_1}$ ~
compatible with the requirement ~$M_{h^0}>50$ GeV ~is attained, becomes equal
to ~$M_{\tilde t_L}$.  ~To the left of the maximum  ~
$M_{\tilde t_R}^{min} > M_{\tilde t_L}$ ~and ~$M_{\tilde t_1}$ ~grows
with ~$M_{\tilde t_L}$ ~(which is related to ~$M_{\tilde b_L}$). ~For larger
values of ~$\tan\beta$ ~we always have ~$M_{\tilde t_R}^{min}<M_{\tilde t_L}$ ~
and no maxima appear.

On the top of the constraints on the lighter stop mass following from the
requirement ~$M_{h^0}>50$ GeV ~come the constraints from ~$M_W$ ~measurement.
However, before we study the impact of the latter, we would like to
improve the calculation of ~$\Delta r$ ~done in ref. \cite{MY_DR} including
corrections to the Higgs boson masses.
In refs. \cite{MY_DR,DS} the effects of these correction
on ~$\Delta r$ ~have not been taken into account. However, although they
are formally a two--loop effects, they should be included, especially for
low values of ~$\tan\beta$ ~for which the corrections to the lighter
Higgs boson mass can reach even 80 GeV \cite{HJEBM} and exceed many times
the tree level mass itself
\footnote{The inclusion of these corrections does not affect the overall
          limits on ~$\Delta r$ ~(for fixed ~$m_t$) given in \cite{MY_DR}
          because both, the lower and the upper bounds are attained for large ~
          $\tan\beta\gg10$ ~for which the corrections have been estimate
          to be small -- $\delta(\Delta r)\sim 0.001$.}.
Most easily they can be taken into account in the effective potential
approach. For given values of the top and stop masses, using the formulae
given in ref. \cite{EZ} we calculate the corrected values of ~$M_{h^0}$ ~and ~
$M_{H^0}$ ~as well as the corrected mixing angle ~$\alpha$. These corrected
quantities are subsequently used in formulae for the gauge boson self--energies
as given e.g. in \cite{MY_H} instead of the tree level ones. Such a procedure
is consistent as far as gauge boson self--energies are concerned and does not
spoil the on--shell renormalization scheme.

The dependence of ~$\Delta r$ ~on the MSSM Higgs sector parameters ~$M_{A^0}$
and ~$\tan\beta$ ~with (solid lines) and without (dotted lines) radiative
corrections to the scalar Higgs boson masses and couplings included
is ilustrated in Fig.2. In Fig. 2a ~$m_t = 175$ GeV ~and ~
$M_{\tilde t_1}\sim M_{\tilde t_2}\sim 1000$ GeV ~so that  the
corrections to ~$M_{H^0,h^0}$ ~are positive. In Fig. 2b ~$m_t=160$ GeV and ~
$M_{\tilde t_1}=60$ ~and $M_{\tilde t_2}\sim150$ GeV leading to negative
mass corrections. In both cases the  {\sl left-right}
mixing vanishes and masses of the charginos, neutralinos, sleptons and
squarks from the first two generations as well as the mass of the righ
handed sbottom are taken of order ~${\cal O}$(1 TeV).
As can be seen, the corrections are more important for the case with
heavy top squarks and small values of the ~$\tan\beta$.
In these plots we do not care about ~$h^0$ ~mass apart from requiring it to
be greater than zero. This eliminates the ~$M_{A^0}<185$ GeV ~part
of the curve for ~$\tan\beta=1.5$ ~in Fig. 2b.
Fig. 2a confirms the estimates given in \cite{MY_DR} that for large ~
$\tan\beta$ ~(for which in MSSM the upper bound of ~$\Delta r$ ~is reached)
the corrections to ~$\Delta r$ ~are of order 0.001.
Notice, that the corrections do not change the fact that for fixed ~
$\tan\beta$ ~maximum of ~$\Delta r$ ~is reached in the limit of large values
of ~$M_{A^0}$. ~Since in the same limit the mass of ~$h^0$ ~is maximized for
fixed ~$\tan\beta$, ~in order to derive absolute bounds on the top
squark masses, we will keep ~$M_{A^0} = 500$ GeV. ~For the same reason
we will always keep charginos, neutralinos, sleptons as well as squarks
from the two first generations of order ~${\cal O}$(1 TeV).

We turn now to the genuine stop/sbottom sector contribution to ~$\Delta r$. ~In
the limit of no {\sl left-right} mixing the influence of the ~$\tilde t_R$ ~
mass on ~$\Delta r$ ~is only indirect, through corrections to the Higgs boson
masses since purely right-handed sfermions decouple almost completely
(apart from their very small contribution through ~$\Delta\alpha_{EM}$, ~see
\cite{MY_DR}) ~from ~$\Delta r$.  ~In this
limit ~$\Delta r$ ~receives negative contribution from ~$\tilde t_L -
\tilde b_L$ ~mass splitting which leads, for large ~$m_t$ ~and
light ~$\tilde b_L$, ~to unacceptably  low values of ~$\Delta r$. ~
More precisely, for example for ~$\tan\beta=2$ ~and ~
$M_{\tilde b_L}=$120 (150) GeV ($M_{A^0}=500$ GeV ~and
other SUSY particles very heavy) ~$\Delta r$ ~becomes
smaller than ~0.034 ~for ~$m_t=$154 (158) GeV ~and smaller than ~0.024 ~for ~
$m_t=$173 (178) GeV. ~For
given ~$m_t$, ~~$\Delta r$ ~can be increased by taking ~$\tilde b_L$ ~and ~
$\tilde t_L$ ~squarks heavy enough and/or switching on the {\sl left-right}
mixing which makes the negative contribution to ~$\Delta r$ ~from
stop/sbottom sector smaller in absolute magnitude \cite{MY_DR}.
The typical features of the squarks contribution to ~$\Delta r$ ~are
illustrated in Figs. 3a-d where for ~$\tan\beta=2$ ~and ~$m_t = 174$ GeV  ~we
plot ~$\Delta r$ ~as a function of the lighter
stop mass for four different values of ~$M_{\tilde b_L}$. ~At the rightmost
point of each curve the {\sl left-right} mixing vanishes (i.e. ~$A_t = 0$
there) and the corresponding mass of ~$M_{\tilde t_R}$ ~is used to label
different  curves. The {\sl left-right} mixing (i.e. the parameter ~$A_t$)
increases along the curves from the right to the left.
If there were no corrections to the Higgs boson masses each curve would extend
up to ~$M_{\tilde t_1}=45$ GeV -- the current direct experimental LEP limit.
However, for too low values of ~$M_{\tilde t_1}$ ~and large {\sl left-right}
mixing in the stop sector, the negative corrections make ~$M_{h^0} < 50$ GeV
and the curves are cutted off at the value of ~$A_t$ ~for which ~
$M_{h^0}$ ~reaches 50 GeV. ~The increase
of ~$\Delta r$ ~with increasing ~$M_{\tilde t_R}$ ~at the
rightmost points of the curves is also due to inclusion of radiative
corrections to the Higgs sector.

As can be seen in Fig.3, ~for low values of ~$M_{\tilde b_L}$ ~imposing the
constraint ~$\Delta r > 0.024$ ~eliminates completely curves for
small ~$M_{\tilde t_R}$ ~strenghtening therefore the lower bound
on ~$M_{\tilde t_1}$. ~For
given ~$M_{\tilde b_L}$ ~and ~$M_{\tilde t_R} > M_{\tilde t_L}$ ~the
lighter stop mass is always bounded from above by ~$M_{\tilde t_L}$ ~itself
because it is ~$\tilde t_L$ ~which plays the role of ~$\tilde t_1$  ~in
the limit of vanishing {\sl left-right} mixing. For ~
$M_{\tilde b_L} <$130 GeV ~however, this upper bound on ~$M_{\tilde t_1}$ ~is
shifted to the lower values by the
requirement ~$\Delta r > 0.024$.  ~The magnitude of this shift decreases
with increasing ~$M_{\tilde t_R}$. ~Therefore, the overall upper bound
on ~$M_{\tilde t_1}$ ~(for fixed ~$M_{\tilde b_L}$) ~depends on how big
hierarchy ~$M_{\tilde b_L} \ll M_{\tilde t_R}$ ~is allowed by the naturality
criterion. In order not to be too restrictive when producing the overall
bounds on ~$M_{\tilde t_1}$ ~we took 1 TeV as an upper limit for ~
$M_{\tilde t_R}$.

Scanning over values of ~$M_{\tilde t_R}$ ~and ~$A_t$ ~for
given ~$M_{\tilde b_L}$ ~one arrives at the absolute  bounds
on ~$M_{\tilde t_1}$. ~These bounds are demonstarted
in Fig. 4a  for ~$m_t=$174 GeV ~(and ~$2\sigma$ ~constraints imposed
on ~$\Delta r$) ~and in Fig. 4b for ~
$m_t=$160 GeV ~(with ~$\Delta r$ ~restricted to ~$1\sigma$) ~for few
values of ~$\tan\beta$. ~As has been discussed above, for low values of ~
$M_{\tilde b_L}$ ~(close to 120 GeV) the upper bound on ~$M_{\tilde t_1}$ ~
comes from the lower ~(2 or 1$\sigma$) ~bound on ~$\Delta r$. ~This effect is
not very big for ~$m_t=174$ GeV -- for ~$\tan\beta=2 (1.5)$ ~
$M_{\tilde t_1}^{max}$ ~is lower than ~$M_{\tilde t_L}$ ~by 5 (15) GeV ~for ~
$M_{\tilde b_L}=120$ GeV ~and dissapears for ~
$M_{\tilde b_L}>130$ (140) GeV -- but is more pronounced for ~
$m_t = 160$ GeV ~where for ~$\tan\beta= 2$ (2.5) ~$M_{\tilde t_1}^{max}$ ~is
lower than ~$M_{\tilde t_L}$ ~by 27 (20) GeV ~for ~$M_{\tilde b_L}=120$ GeV
and these effect extends up to ~$M_{\tilde b_L} = 170$ (155) GeV.

The lower limit for ~$\tan\beta = 1.5$ ~comes solely from the condition ~
$M_{h^0}>50$ GeV ~(and is therefore given by the same lines as in Fig.1). ~
However,  for ~$\tan\beta > 1.75$ ~due to the restriction iposed on ~
$\Delta r$ ~the lower bound on ~$M_{\tilde t_1}$ ~is stronger
than it follows from the requirement ~$M_{h^0}>50$ GeV ~only. It is
also interesting to notice that for ~$m_t=160$ GeV ~and ~$\tan\beta=1.5$
(1.75) ~the combined constraints from ~$M_{h^0}$ ~and ~$\Delta r$ ~eliminate
values of ~$M_{\tilde b_L}$ ~up to 155 (130) GeV.

To conclude, taking into account constraint coming from ~$M_W$ ~measurement
by UA2, CDF and D0
and from the unsuccesful MSSM Higgs boson search at LEP1, we have obtained
rather strong (for small values of ~$\tan\beta$) ~constraints on the allowed
regions in the ~($M_{\tilde b_L}, M_{\tilde t_1}$) ~plane. Let us stress,
that since we have kept all other SUSY particles very heavy, the obtained
allowed regions are maximal. Making e.g. sneutrinos light would result
in much stronger constraints on the ~($M_{\tilde b_L}, M_{\tilde t_1}$) ~
plane because of the additional negative contributions to ~$\Delta r$. ~On
the other hand it is hard to imagine that sleptons are much
heavier than squarks (actually, all the renormalization group study of the
MSSM embeded into supergravity predict the opposite). Therefore, our bounds
are certainly the weakest ones.

\vskip 0.5cm
\noindent {\bf Acknowledgments.} The author would like to thank Professor
W. Hollik for kind hospitality extended to him at the University of Karlsruhe
where this work has been completed and for several discussions. He would
also like to thank Professor S. Pokorski for his interest in this work
and several comments and Professor Gennadij G. Volkov for inspiring question.

\newpage
\noindent {\bf FIGURE CAPTIONS}
\vskip 0.5cm

\noindent {\bf Figure 1.}

\noindent Lower bounds on the
lighter top squark mass from ~$M_{h^0}>$50 GeV ~for
different values of the ~$\tan\beta$ ~(marked on the curves)
as a function of ~$M_{\tilde b_L}$ - the mass of the left handed sbottom
and two different masses of the top quark.
\vskip 0.3cm

\noindent {\bf Figure 2.}

\noindent Dependence of ~$\Delta r$ ~on ~$M_{A^0}$ ~for different values
of ~$\tan\beta$. ~Solid (dotted) lines show the results with corrections to
the Higgs boson masses included (neglected).

a) ~$m_t=174$ GeV ~and ~$M_{\tilde t_1}\sim M_{\tilde t_2}\sim 1000$ GeV.

b) ~$m_t=160$ GeV ~and
{}~$M_{\tilde t_1}=60$ ~and ~$M_{\tilde t_2}\sim150$ GeV.

\noindent All other SUSY particles are heavy, ~${\cal O}$(1 TeV).
\vskip 0.3cm

\noindent {\bf Figure 3.}

\noindent $\Delta r$ ~as a function of the lighter stop
mass, ~$M_{\tilde t_1}$ ~for ~$m_t = 174$ GeV, ~$\tan\beta =2$ ~and four
different masses of
the left handed sbottom ~$M_{\tilde b_L}$. ~Numbers on the curves are
the corresponding values of ~$M_{\tilde t_R}$. ~The ~{\sl left-right}
mixing, i.e. ~$A_t$, ~vanishes at the rightmost point of each curve
and increases from the right to the left. The curves are cutted off
at the values of ~$A_t$ ~for which ~$M_{h^0}$ ~becomes lighter than 50 GeV.
The ~$2\sigma$ ~lower limit on ~$\Delta r$ ~is marked by dotted lines.
\vskip 0.3cm

\noindent {\bf Figure 4.}

\noindent Allowed regions in
the~ ($M_{\tilde b_L}, M_{\tilde t_1}$) ~plane for different values
of ~$\tan\beta$. ~For ~$m_t=$174 GeV ~in  Fig. 4a  ~$2\sigma$ ~bound
on ~$\Delta r$ ~is imposed. For ~$m_t=$160 GeV ~in  Fig. 4b  ~$\Delta r$ ~is
restricted to its ~$1\sigma$ ~experimental value.
Allowed regions for ~$\tan\beta=$ 1.5, 1.75, 2 and 2.5 ~are marked by
dashed, dot-dashed, solid and dotted lines respectively.
\end{document}